\documentclass[twocolumn,journal]{IEEEtran}

\usepackage{amsfonts, amssymb, amsmath, cite, enumerate, amsthm, bm, caption, subfig, psfrag, cases, mathtools}
\usepackage[usenames,dvipsnames]{color}
\ifCLASSINFOpdf
   \usepackage[pdftex]{graphicx}
   \DeclareGraphicsExtensions{.pdf,.jpeg,.png}
\else
   \usepackage[dvips]{graphicx}
   \DeclareGraphicsExtensions{.eps}
\fi

\DeclareMathOperator{\E}{\mathbb{E}}

\DeclareMathOperator*{\tr}{tr}

\newcommand{\ip}[2]{\left\langle#1,#2\right\rangle}
\newcommand{\norm}[1]{\left\lVert#1\right\rVert}
\renewcommand{\(}{\left(}
\renewcommand{\)}{\right)}
\renewcommand{\[}{\left[}
\renewcommand{\]}{\right]}

\def \C {\mathbb{C}}

\def \P {\mathbb{P}}
\def \R {\mathbb{R}}

\def \S {\mathbb{S}}

\def \e {\varepsilon}

\def \< {\langle}
\def \> {\rangle}

\def \vx {\bm{x}}
\def \vu {\bm{u}}
\def \vv {\bm{v}}
\def \vy {\bm{y}}
\def \vz {\bm{z}}
\def \vX {\bm{X}}
\def \vY {\bm{Y}}
\def \vZ {\bm{Z}}
\def \vA {\bm{A}}
\def \vB {\bm{B}}
\def \vC {\bm{C}}
\def \vD {\bm{D}}

\def \vW {\bm{W}}

\def \vI {\bm{I}}

\def \vP {\bm{P}}

\def \vT {\bm{T}}

\def \vSigma {\bm{\Sigma}}

\def \vLambda {\bm{\Lambda}}

\theoremstyle{plain}
\newtheorem{theorem}{Theorem}
\newtheorem{proposition}{Proposition}
\newtheorem{corollary}{Corollary}
\newtheorem{definition}{Definition}

\newtheorem{lemma}{Lemma}
\newtheorem{fact}{Fact}

\theoremstyle{remark}
\newtheorem{remark}{Remark}
\newtheorem{example}{Example}

\begin{document}
\title{Covariance Matrix Estimation from Correlated Sub-Gaussian Samples}

\author{Xu~Zhang, Wei~Cui, and Yulong~Liu
\thanks{X.~Zhang and W.~Cui are with the School of Information and Electronics, Beijing Institute of Technology, Beijing 100081, China (e-mail: connorzx@bit.edu.cn;cuiwei@bit.edu.cn).}
\thanks{Y.~Liu is with the School of Physics, Beijing Institute of Technology, Beijing 100081, China (e-mail: yulongliu@bit.edu.cn).}
}

\maketitle

 \begin{abstract}
 This paper studies the problem of estimating a covariance matrix from correlated sub-Gaussian samples. We consider using the correlated sample covariance matrix estimator to approximate the true covariance matrix. We establish non-asymptotic error bounds for this estimator in both real and complex cases. Our theoretical results show that the error bounds are determined by the signal dimension $n$, the sample size $m$ and the correlation pattern $\vB$. In particular, when the correlation pattern $\vB$ satisfies $\tr(\vB)=m$, $||\vB||_{F}=O(m^{1/2})$, and $||\vB||=O(1)$, these results reveal that $O(n)$ samples are sufficient to accurately estimate the covariance matrix from correlated sub-Gaussian samples. Numerical simulations are presented to show the correctness of the theoretical results.

 \end{abstract}


%
\IEEEpeerreviewmaketitle

\section{Introduction}
Covariance matrix estimation is concerned with the problem of estimating the covariance matrix from a collection of samples, which is a basic problem in modern multivariate analysis and arises in diverse fields such as signal processing \cite{krim1996two}, machine learning \cite{friedman2001elements}, statistics \cite{cai2016estimating}, and finance \cite{fan2016overview}. Typical applications in signal processing include Capon's estimator \cite{capon1969high}, MUltiple SIgnal Classification (MUSIC) \cite{schmidt1986multiple}, Estimation of Signal Parameter via Rotation Invariance Techniques (ESPRIT) \cite{roy1989esprit}, and their variants \cite{krim1996two}.

Consider a centered random vector $\vx\in \R^n$ with the covariance matrix $\vSigma=\E [\vx \vx^T]$, where $\vSigma$ is an $n \times n$ positive definite matrix. Let $\vx_1,\ldots,\vx_m$ be independent copies of $\vx$. A classical unbiased estimator for $\vSigma$ is the sample covariance matrix
\begin{equation*}
\widetilde{\vSigma}= \frac{1}{m} \sum \limits_{k=1}^m \vx_k \vx_k^T = \frac{1}{m}\vX \vX^T,
\end{equation*}
where $\vX=[\vx_1,\ldots,\vx_m]\in \R^{n \times m}$.
A basic question is to determine the minimal sample size $m$ which guarantees that $\vSigma$ is accurately estimated by $\widetilde{\vSigma}$. The past few decades have witnessed great interest in different instances of this question \cite{bai1993limit,aubrun2007sampling,vershynin2010into,adamczak2010quantitative,adamczak2011sharp,vershynin2012close,srivastava2013covariance,koltchinskii2017concentration}. For example, Vershynin \cite{vershynin2010into} establishes that $m=O(n)$ samples are enough for independent sub-Gaussian samples, where $O(n)$ means that the required samples is a linear function of the signal dimension $n$; Vershynin \cite{vershynin2012close} also shows that $O(n\log n)$ samples are sufficient for independent heavy tailed samples; and Srivastava and Vershynin \cite{srivastava2013covariance} illustrate that $O(n)$ is the optimal bound for independent samples which are sampled from log-concave distributions.

In many practical applications, however, we often have access to correlated signal samples rather than independent samples.  A typical example in signal processing is that the received samples are often correlated when the signals are transmitted in multipath channel \cite{ramirez2010detection,huang2013detection} or the signal sources interfere with each other \cite{shiu2000fading,liu2007training}. Another important instance in portfolio management and risk assessment is that the returns between different assets are correlated on short time scales, i.e., the Epps effect \cite{epps1979comovements,munnix2010impact}. A basic problem in these scenarios is how many correlated samples are required to have a good estimation of the true covariance matrix?

In a recent paper \cite{cui2019covariance}, the present authors consider covariance matrix estimation from linearly-correlated Gaussian samples. More precisely, let $\vx_1,\ldots,\vx_m \in \R^n$ be independent and identically distributed (i.i.d.) Gaussian vectors with zero mean and covariance matrix $\vSigma$. Assume that we observe $m$ linearly-correlated samples $\{\bm{y}_k\}_{k=1}^{m}$, i.e.,
\begin{equation} \label{Linearly-correlatedModel}
 \vY=\vX \bm{\Lambda},
\end{equation}
where $\bm{Y}=[\vy_1,\ldots,\vy_m]$, $\vX=[\vx_1,\ldots,\vx_m]$, and $\vLambda \in \R^{m \times m}$  is an arbitrary matrix. A natural estimator for $\vSigma$ in the correlated case is the following correlated sample covariance matrix (see, e.g., \cite{collins2013compound, burda2011applying, chuah2002capacity})
\begin{equation} \label{eq: SCM1}
	\hat{\vSigma}=\frac{1}{m} \sum \limits_{k=1}^m \vy_k \vy_k^T = \frac{1}{m}\vX \vLambda \vLambda^T \vX^T = \frac{1}{m}\vX \vB \vX^T.
\end{equation}
The theoretical results in \cite{cui2019covariance} establish that the approximation error by $\hat{\vSigma}$ is determined by the signal dimension $n$, the sample size $m$, and the shape parameter $\vB = \vLambda \vLambda^T$ of the correlated sample covariance matrix. In particular, if the shape parameter is a class of important Toeplitz matrices, where $\vB$ satisfies $\tr(\vB)=m$, $||\vB||_{F}=O(m^{1/2})$, and $||\vB||=O(1)$, these results reveal that $m=O(n)$ samples are also sufficient for linearly-correlated Gaussian samples.

In the current paper, we generalize our previous work \cite{cui2019covariance} in three important aspects:
\begin{itemize}
  \item From symmetric $\vB$ to general $\vB$: In the linearly-correlated model \eqref{Linearly-correlatedModel}, the shape parameter $\vB = \vLambda \vLambda^T$ is obviously symmetric (and even positive semi-definite). However, in some applications, the shape parameter $\vB$ might be nonsymmetric, which allows more general correlated patterns and makes our previous theory for symmetric $\vB$ inapplicable. For instance, when investigating the group symmetric properties of sample covariance matrices, the shape matrix $\vB$ is a class of skew-symmetric matrices \cite{shah2012group, soloveychik2014error}. This fact motivates us to develop new theoretical results for general $\vB$.

  \item From Gaussian samples to sub-Gaussian samples: This extension enables our theoretical results applicable for larger classes of random samples, such as Gaussian, Bernoulli and any bounded random samples.

  \item From real samples to complex samples: This generalization is natural since complex samples are ubiquitous in signal processing applications.
\end{itemize}

Under the above generalized settings, we develop a totally new strategy to establish a non-asymptotic analysis for covariance matrix estimation from correlated sub-Gaussian samples. Our results show that the error bounds are also determined by the signal dimension $n$, the sample size $m$, and the shape parameter $\vB$. Particularly, $O(n)$ samples are sufficient to estimate the covariance matrix accurately from correlated sub-Gaussian samples, provided that the correlation pattern $\vB$ satisfies $\tr(\vB)=m$, $||\vB||_{F}=O(m^{1/2})$, and $||\vB||=O(1)$, which shares the same order of sample size as covariance matrix estimation from correlated linearly-correlated Gaussian samples.

This paper is organized as follows. Preliminaries are provided in Section \ref{sec: preliminaries}. Concentration inequalities of the general compound Wishart matrix are established in Section \ref{sec: Concentration Inequality}. The performance analysis of covariance matrix estimation from correlated sub-Gaussian samples is presented in Section \ref{sec: CECC}. Simulations are provided in Section \ref{sec: Simulation}. Conclusions and future works are given in Section \ref{sec: Conclusion}.

The following notation is adopted in the paper: $\R$ denotes the real domain while $\C$ denotes the complex domain. ${\rm{Re}(\cdot)}$ returns the real part and  ${\rm {Im}}(\cdot)$ returns the imaginary part of a scalar, vector or matrix. Lowercase letters are reserved for scalars, e.g., $x,y,z$; lowercase boldface letters are used for vectors, e.g., $\vx,\vy,\vz$; and uppercase boldface letters are applied for matrices, e.g., $\vX,\vY,\vZ$.
For a vector $\vx$, $x_i$ is the $i$-th component of $\vx$. For a matrix $\vX$, $X_{ij}$ denotes the $(i,j)$-th entry of the matrix. $\vI_n$ is the $n$-dimensional identity matrix. $(\cdot)^T$ returns the transpose and $(\cdot)^H$ returns the conjugate transpose. The $\ell_p$ norm of a vector $\vx$ is denoted by $\norm{\vx}_p=(\sum_{i=1}^{n}|x_i|^p)^{1/p}$. The $L^p$ norm of a random variable $x$ is defined as $\norm{x}_{L^p}=\(\E|x|^p\)^{1/p}$. $\norm{\cdot}_{F}$ denotes the Frobenius norm, $\norm{\cdot}$ denotes the spectral norm, and $\ip{\cdot}{\cdot}$ denotes the inner product.
$\mathbb{S}^{n-1}$ denotes the unit sphere in $n$-dimensional real or complex space under $\ell_2$-norm. $O(n)$ means the order of the growth is a linear function of $n$.
The notations $c,\,C, C', $ and $C''$ are absolute positive constants which may vary with different cases.

\section{preliminaries} \label{sec: preliminaries}
In this section, we review some related definitions and facts, which will be used in this paper.
\subsection{Some definitions}
We begin by introducing some definitions from high dimensional probability theory.
\begin{definition}[Sub-Gaussian random variables] A random variable $x$ is a \emph{sub-Gaussian random variable} if the Orlicz norm
	\begin{equation}\label{SubGaussianDef}
	  \norm{x}_{\psi_2}=\inf \left\{t>0: \E \exp\(\frac{x^2}{t^2}\)\le 2\right\}
	\end{equation}
is finite. The sub-Gaussian norm of $x$, denoted $\norm{x}_{\psi_2}$, is defined to be the smallest $t$ in \eqref{SubGaussianDef}.	
\end{definition}
There are several equivalent definitions used in the literature, see e.g., \cite[Proposition 2.7.1]{vershynin2017high}. Important examples of sub-Gaussian random variables include Gaussian, Bernoulli and all bounded random variables.

\begin{definition}[Sub-Gaussian random vectors]
	A random vector $\vx \in \R^n$ is called a \emph{sub-Gaussian random vector} if all of its one-dimensional marginals are sub-Gaussian, and its sub-Gaussian norm is defined as
	$$
	\norm{\vx}_{\psi_2} = \sup_{\vy \in \S^{n-1}} \norm{\left\langle \vx,\vy \right\rangle}_{\psi_2}.
	$$
\end{definition}

\begin{definition}[Isotropic vectors]
A random vector $\vx \in \R^n$  is called \emph{isotropic} if it satisfies $\E \vx \vx^T = \bm{I}_n$.
\end{definition}

Clearly, for any random vector $\vx$ with positive definite covariance matrix $\E [\vx\vx^T]=\vSigma$, then $\vy=\vSigma^{-{1}/{2}}\vx$ is an isotropic vector.

We say that an $n \times n$ random matrix $\vW$ is a compound Wishart matrix with shape parameter $\vB$ and scale parameter $\vSigma$ if $\vW = \frac{1}{m}\vX \vB \vX^T$, where $\vX = [\vx_1,\ldots,\vx_m]$, $\vx_1,\ldots,\vx_m \sim \mathcal{N}(\bm{0},\vSigma)$ are independent Gaussian vectors, and $\vB$ is an arbitrary real $m \times m$ matrix \cite{speicher1998combinatorial}. The following definition directly extends compound Wishart matrices for Gaussian distribution to the sub-Gaussian case.
\begin{definition}[General compound Wishart matrices] \label{def: Compound Wishart matrices}
    Let $\vx_1,\ldots,\vx_m$ be i.i.d. sub-Gaussian random vectors with zero mean and covariance matrix $\vSigma$, and let $\vB \in \R^{m \times m}$ be an arbitrary matrix. The matrix $\vW \in \R^{n \times n}$ is called a \emph{general compound Wishart matrix} with shape parameter $\vB$ and scale parameter $\vSigma$ if $\vW$ has the following form $$\vW = \frac{1}{m}\vX \vB \vX^T,$$ where $\vX = [\vx_1,\ldots,\vx_m]$.
\end{definition}

\subsection{Some useful facts}
We introduce some useful facts which will be used to derive our main results.

Recall that let $\mathcal{K} \subset \R^n$ and $\e>0$. A subset $\mathcal{N} \subset \mathcal{K}$ is called an $\e$-net of $\mathcal{K}$ if
	$$\forall~\vx \in \mathcal{K}, ~~\exists~\vx_0 \in \mathcal{N} ~\text{such that}~ \norm{\vx-\vx_0}_2 \le \e.$$

\begin{fact}[Exercise 4.4.3 and Corollary 4.2.13, \cite{vershynin2017high}] \label{lm: covering} Let $\vA$ be an $m \times n$ real matrix and $\e \in [0,1/2)$. Let $\mathcal{N}$  be an $\e$-net of the unit sphere $\mathbb{S}^{n-1}$ and $\mathcal{M}$  be an $\e$-net of the unit sphere $\mathbb{S}^{m-1}$. Then we have
$$
\norm{\vA} \le \frac{1}{1-2\e} \sup \limits_{\vx \in \mathcal{N},\vy \in \mathcal{M}} \ip{\vA \vx}{\vy}.
$$
	Furthermore, there exist $\e$-nets $\mathcal{N}$ and  $\mathcal{M}$ with cardinalities
	$$
	|\mathcal{N}| \le \(1+\frac{2}{\e}\)^n \text{~and~} 	|\mathcal{M}| \le \(1+\frac{2}{\e}\)^m.
	$$
\end{fact}

\begin{fact}[Hanson-Wright inequality, Theorem 1.1, \cite{rudelson2013hanson}] \label{lm: HW}
Let $\vx \in \R^m$ be a sub-Gaussian vector whose entries are independent centered sub-Gaussian variables with $\norm{x_i}_{\psi_2}\le K, i=1,\ldots,m$. Let $\vB \in \R^{m \times m}$ be a fixed matrix. Then for any $t \geq 0$, we have
\begin{multline*}
	\P\(|\ip{\vB \vx}{\vx} - \E\ip{\vB \vx}{\vx}| \ge t\) \\
	\le 2 \exp\[-c \min\(\frac{t^2}{K^4 \norm{\vB}_F^2},\frac{t}{K^2\norm{\vB}}\) \].
\end{multline*}
\end{fact}


\subsection{Related results} \label{sec: problem formulation}

To aid comparisons, we review some highly related results in the literature. For independent sub-Gaussian samples, Proposition \ref{pp: sg} indicates that $O(n)$ samples is sufficient to obtain an accurate estimation of the covariance matrix. In the linear-correlated Gaussian model, Proposition \ref{lm: wishart1} shows that if the correlation parameter $\vB$ satisfies $\tr(\vB)=m$, $||\vB||_{F}=O(m^{1/2})$, and $||\vB||=O(1)$, then $O(n)$ samples are enough to approximate the covariance matrix well.


\begin{proposition}[Theorem 4.7.1, \cite{vershynin2017high}] \label{pp: sg}
 Let $\vx \in \R^n$ be a centered sub-Gaussian vector with the positive definite covariance matrix $\vSigma=\E[\vx \vx^T]$. Let $\vx_1,\ldots,\vx_m \in \R^n$ be independent copies of $\vx$ and $\vX=[\vx_1,\ldots,\vx_m]$. Suppose that there exists $K\ge 1$ such that
 \begin{equation} \label{eq: psi_2_norm_condition}
 	\norm{\ip{\vx}{\vy}}_{\psi_2} \le K \norm{\ip{\vx}{\vy}}_{L^2}, \text{~for any~} \vy \in \R^n.
 \end{equation}
 Then for any $\delta \geq 0$, the sample covariance matrix $\widetilde{\vSigma}=\frac{1}{m}\vX \vX^T$ satisfies
 \begin{equation*}
 	||\widetilde{\vSigma}-\vSigma|| \le CK^2 \(\sqrt{\frac{n+\delta}{m}}+\frac{n+\delta}{m}\)\norm{\vSigma}
 \end{equation*}
 with probability as least $1-2\exp(-\delta)$. Furthermore,
$$\E||\widetilde{\vSigma}-\vSigma|| \le CK^2 \( \sqrt{\frac{n}{m}}+\frac{n}{m}\) \norm{\vSigma}.$$
\end{proposition}
\begin{remark} Note that the condition \eqref{eq: psi_2_norm_condition} essentially provides an upper bound for the sub-Gaussian norm of $\vx$. To see this, let $\vSigma=\vI$, i.e., the random vector $\vx$ is isotropic, then the condition \eqref{eq: psi_2_norm_condition} becomes
$$
 \norm{\ip{\vx}{\vy}}_{\psi_2} \le K \norm{\ip{\vx}{\vy}}_{L^2} = K\|\vy\|_2, \text{~for any~} \vy \in \R^n,
$$
and hence
\begin{align*}
	\norm{\vx}_{\psi_2}=\sup \limits_{\vy \in \S^{n-1}} \norm{\left\langle \vx,\vy \right\rangle}_{\psi_2}\le K.
\end{align*}
\end{remark}

\begin{proposition}[Theorem 2,\cite{cui2019covariance}] \label{lm: wishart1}
	Let $\vx_1,\ldots,\vx_m$ be independent Gaussian vectors with zero mean and covariance matrix $\vSigma \in \R^{n \times n}$, where $\vSigma$ is a positive definite matrix. Let the correlated sample covariance matrix  estimator be $\hat{\vSigma} = \frac{1}{m}\vX \vB \vX^T,$ where $\vB \in \R ^{m \times m}$ is a symmetric matrix and $\vX=[\vx_1,\ldots,\vx_m]$.
	{Then for any $\delta \geq 0$, the event
		\begin{multline*}
		\norm{\hat{\vSigma}-\vSigma} \le \left| \frac{\tr(\vB)}{m}-1 \right|  ||\vSigma||
		\\ + C \(\frac{\sqrt{n+\delta} \norm{\vB}_F + (n+\delta) \norm{\vB}}{m}\)\norm{\vSigma}
		\end{multline*}
		holds with probability at least $1-2\exp(-\delta)$.} Furthermore,
	\begin{multline*}
	\E \norm{\hat{\vSigma}-\vSigma} \le  \left| \frac{\tr(\vB)}{m}-1 \right|  ||\vSigma|| \\
	+ C \(\frac{\sqrt{n} \norm{\vB}_F + n \norm{\vB}}{m}\)\norm{\vSigma}.
	\end{multline*}
\end{proposition}

\section{Concentration Inequalities of General Compound Wishart Matrices} \label{sec: Concentration Inequality}

In this section, we establish concentration inequalities for the general compound Wishart matrix in both real and complex cases. These results illustrate that the correlated sample covariance matrix $\hat{\vSigma}= \vX \vB \vX^T/m$ concentrates around it mean $\E \hat{\vSigma}$ with high probability. As we will see in the section \ref{sec: CECC}, these results play a key role in establishing a non-asymmetric analysis for the correlated covariance matrix estimator.

\begin{theorem}[Real case]\label{thm: wishart_SG1}
	Let $\vx_1,\ldots,\vx_m$ be i.i.d. centered sub-Gaussian vectors with the positive definite covariance matrix $\vSigma \in \R^{n \times n}$. Let $\vB \in \R^{m \times m}$ be an arbitrary fixed matrix. Consider the general compound Wishart matrix $\hat{\vSigma}= \vX \vB \vX^T/m$ with $\vX=[\vx_1,\ldots,\vx_m]$. Suppose that there exists $K\ge1$ such that
	\begin{equation} \label{eq: psi_2_norm_condition2}
	\norm{\ip{\vx_i}{\vy}}_{\psi_2} \le K \norm{\ip{\vx_i}{\vy}}_{L^2}, \forall \,  \vy \in \R^n,\, i=1,\ldots,m.
	\end{equation}
	Then for any $\delta \geq 0$, the following event
	\begin{multline*}
		\norm{\hat{\vSigma}-\E\hat{\vSigma}} \\ \le CK^2 \(\frac{\sqrt{n+\delta} \norm{\vB}_F + (n+\delta) \norm{\vB}}{m}\)\norm{\vSigma}
	\end{multline*}
	holds with probability at least $1-2\exp(-\delta)$.
	Furthermore,
	\begin{equation} \label{thm1: expectationbound}
	\E\norm{\hat{\vSigma}-\E\hat{\vSigma} } \le CK^2 \(\frac{ \sqrt{n} \norm{\vB}_F + n \norm{\vB} }{m}\)\norm{\vSigma}.
	\end{equation}
\end{theorem}
\begin{IEEEproof}
	See {Appendix \ref{appdix: Proof_Thm1}}.
\end{IEEEproof}

\begin{remark}
Theorem \ref{lm: wishart1} illustrates that the error bounds depend on the signal dimension $n$, the sample size $m$, and the shape parameter $\vB$. In particular, if $||\vB||_{F}=O(m^{1/2})$ and $||\vB||=O(1)$, then this result reveals that $m = O(n)$ samples are sufficient to approximate the general compound Wishart matrix $\hat{\vSigma}$ (by its expectation $\E\hat{\vSigma}$) accurately.
\end{remark}

\begin{remark}
It should be pointed out that the proof of Theorem \ref{thm: wishart_SG1} requires a totally new strategy in contrast to the linear-correlated Gaussian model in \cite{cui2019covariance}. This is because many useful properties in the linear-correlated Gaussian model (e.g., the rotation invariance property of Gaussian distribution and symmetry of the shape parameter B) are non-available in the generalized case.
\end{remark}

\begin{remark}[Related works for general $\vB$]

In \cite{soloveychik2014error}, Soloveychik establishes the following expectation bound for the Gaussian samples
\begin{multline*}
  \E \norm{\hat{\vSigma}-\E\hat{\vSigma}}  \\ \le \frac{24 \lceil\log 2n\rceil^2 \sqrt{n} (4\|\vB\| + \sqrt{\pi} \|\vB\|_{F}/\|\vB\|)}{m} \norm{\vSigma},
\end{multline*}
which implies that if $||\vB||_{F}=O(m^{1/2})$ and $||\vB||=O(1)$, then $m = O(n\log^4n)$ samples are { sufficient} to approximate the compound Wishart matrix $\hat{\vSigma}$ accurately.

In \cite{paulin2016efron}, Paulin et al. establish the concentration of $\hat{\vSigma}$ in both expectation and tail forms for the bounded samples (i.e., each entry of $\vX$ is bounded by an absolute positive constant $L$). The expectation bound in \cite{paulin2016efron} is
\begin{equation*}\label{leq: Paulin}
	\E ||\hat{\vSigma}- \E\hat{\vSigma}||  \le  \frac{ 2 \sqrt{ v(\vB) \log n } + 32\sqrt{3} L n \log n ||\bm{B}||}{m},
\end{equation*}
where $v(\vB)=44(n \sigma^2+L^2)\|\vB\|_{F}^2$ and $\sigma$ is the standard deviation of each entry of $\vX$. It is not hard to find that if $||\vB||_{F}=O(m^{1/2})$ and $||\vB||=O(1)$, then this bound indicates that $m = O(n\log n)$ samples suffice to approximate the general compound Wishart matrix $\hat{\vSigma}$.

Since Gaussian and bounded random variables belong to sub-Gaussian random variables, the above two results might be regarded as special cases of our result. More importantly, our results improve theirs in the general $\vB$ case. This improvement is critical to obtain the optimal error rate for the covariance matrix estimation from correlated sub-Gaussian samples.

\end{remark}

We then present a complex counterpart of Theorem \ref{thm: wishart_SG1}.

\begin{theorem}[Complex case] \label{thm: wishart_SG2} Consider a complex vector $\vx \in \C^n$ whose real part ${\rm{Re}}(\vx)$ and imaginary part ${\rm{Im}}(\vx)$ are i.i.d. centered sub-Gaussian random vectors. Let  $\vSigma=\E \vx \vx^H$ be the positive definite covariance matrix of $\vx$. Suppose that there exists $K\ge1$ such that
	\begin{equation} \label{eq: psi_2_norm_condition_complex}
	\norm{\ip{\left[
			\begin{array}{c}
			{\rm{Re}} (\vx)       \\
			{\rm{Im}} (\vx)          \\
			\end{array}
			\right]}{\vy}}_{\psi_2} \le 2K \norm{\ip{\left[
			\begin{array}{c}
			{\rm{Re}} (\vx)       \\
			{\rm{Im}} (\vx)          \\
			\end{array}
			\right]}{\vy}}_{L^2}
	\end{equation}
for any $\vy \in \R^{2n}$.
Let  vectors $\vx_1,\ldots,\vx_m \in \C^n$ be independent copies of $\vx$ and $\vB \in \C^{m \times m}$ be a fixed matrix. Consider the general compound Wishart matrix $\hat{\vSigma}= \vX \vB \vX^H/m$ with $\vX=[\vx_1,\ldots,\vx_m]$.
	Then for any $\delta \geq 0$, the following event
	\begin{multline*}
	\norm{\hat{\vSigma}-\E\hat{\vSigma}}  \\\le {CK^2 }\( \frac{\sqrt{n+\delta}\norm{\vB}_F+ (n+\delta) \norm{\vB}}{m}\)  \norm{\vSigma}
	\end{multline*}
	holds with probability at least $1-c\exp(-\delta)$.
	Furthermore,
	\begin{equation*}
	\E\norm{\hat{\vSigma}-\E\hat{\vSigma} } \le {CK^2 } \(\frac{\sqrt{n}\norm{\vB}_F+ n \norm{\vB}}{m}\)  \norm{\vSigma}.
	\end{equation*}
\end{theorem}
\begin{IEEEproof}
	See {Appendix \ref{appdix: Proof_Thm3}}.
\end{IEEEproof}

\section{Covariance Matrix Estimation from Correlated {Sub-Gaussian} Samples} \label{sec: CECC}
In this section, by using concentration inequalities of the general compound Wishart matrix, we establish the non-asymptotic error bounds for the correlated sample covariance matrix estimator in both expectation and tail forms. We also provide some typical examples  to illustrate the theoretical results.

\subsection{Theoretical  guarantees}
\begin{theorem}[Real case] \label{thm: General} Let $\vx_1,\ldots,\vx_m$ be i.i.d. centered sub-Gaussian random vectors with positive definite covariance matrix $\vSigma \in \R^{n \times n}$. Let $\vB \in \R ^{m \times m}$ be an arbitrary matrix. Consider the correlated sample covariance matrix estimator $\hat{\vSigma} = \vX \vB \vX^T/m$  with $\vX=[\vx_1,\ldots,\vx_m]$. Suppose that there exists $K\ge1$ such that
	\begin{equation*}
	\norm{\ip{\vx_i}{\vy}}_{\psi_2} \le K \norm{\ip{\vx_i}{\vy}}_{L^2}, \forall \,  \vy \in \R^n,\, i=1,\ldots,m.
	\end{equation*}
	{Then for any $\delta \geq 0$, the covariance matrix estimator $\hat{\vSigma}$ satisfies
	\begin{multline*}
	\norm{\hat{\vSigma}-\vSigma} \le \left| \frac{\tr(\vB)}{m}-1 \right|  ||\vSigma||
	\\ + CK^2 \(\frac{\sqrt{n+\delta} \norm{\vB}_F + (n+\delta) \norm{\vB}}{m}\)\norm{\vSigma}
	\end{multline*}
	with probability at least $1-2\exp(-\delta)$.} Furthermore,
	\begin{multline*}
	\E \norm{\hat{\vSigma}-\vSigma} \le  \left| \frac{\tr(\vB)}{m}-1 \right|  ||\vSigma|| \\
	+ CK^2 \(\frac{\sqrt{n} \norm{\vB}_F + n  \norm{\vB}}{m}\)\norm{\vSigma}.
	\end{multline*}
\end{theorem}

\begin{IEEEproof} Using the triangle inequality yields
	\begin{equation} \label{mthm: neq_1}
	\E \norm{\hat{\vSigma}-\vSigma}\le \E \norm{\hat{\vSigma}-\E\hat{\vSigma}} +\norm{\E\hat{\vSigma}-{\vSigma}}.
	\end{equation}
	The first term in \eqref{mthm: neq_1} can be easily bounded by using Theorem \ref{thm: wishart_SG1}, i.e.,
	\begin{equation} \label{mthm: neq_2}
	\E \norm{\hat{\vSigma}-\E\hat{\vSigma}} \le CK^2 \(\frac{ \sqrt{n} \norm{\vB}_F + n \norm{\vB} }{m}\)\norm{\vSigma}.
	\end{equation}
    We only need to bound the second term in \eqref{mthm: neq_1}. Since the columns of $\vX$ are centered independent sub-Gaussian vectors, we have
	\begin{align*}
	\E[\hat{\Sigma}_{ij}]&=\frac{1}{m}\sum_{l,k=1}^{m} B_{lk} \E \(X_{il} X_{jk}\)\\
	&=\frac{1}{m}\sum_{l=1}^{m} B_{ll} \E \(X_{il} X_{jl}\)\\
	&=\frac{\tr(\vB)}{m}\Sigma_{ij},
	\end{align*}
	where $X_{ij}$ denotes the $(i,j)$-th entry of the matrix $\vX$, $i=1,\ldots,n, j=1,\ldots,m$. Thus we get
	\begin{equation} \label{mthm: neq_3}
	\E \hat{\vSigma}=\frac{\tr(\vB)}{m} \vSigma.
	\end{equation}
	Substituting \eqref{mthm: neq_2} and \eqref{mthm: neq_3}  into \eqref{mthm: neq_1} yields the expectation bound.

    The tail bound can be obtained by using the following equality
    \begin{align*}
		\P \(\norm{\hat{\vSigma}-\vSigma} \ge t\) &\le \P \(\norm{\hat{\vSigma}-\E\hat{\vSigma}} +  \norm{\E\hat{\vSigma}-{\vSigma}}\ge t\) \\
		& = \P \(\norm{\hat{\vSigma}-\E\hat{\vSigma}} \ge t-\norm{\E\hat{\vSigma}-{\vSigma}}\).
	\end{align*}
	and setting
	\begin{multline*}
	t=t_0 =\norm{\E\hat{\vSigma}-{\vSigma}} \\ +  CK^2 \(\frac{\sqrt{n+\delta} \norm{\vB}_F + (n+\delta) \norm{\vB}}{m}\)\norm{\vSigma}.
	\end{multline*}
	It  follows from Theorem \ref{thm: wishart_SG1} that for any $\delta \geq 0$
	$$
	\P \(\norm{\hat{\vSigma}-\vSigma} \ge t_0\) \le 2 \exp(-\delta),
	$$	
	which completes the proof.
\end{IEEEproof}

\begin{remark} Comparing Theorem \ref{thm: General} and Proposition \ref{lm: wishart1}, it is not hard to find that covariance matrix estimation from the linear-correlated Gaussian samples has the same order of error rate with that from correlated sub-Gaussian samples.
\end{remark}


In particular, if the shape matrix satisfy $\tr(\vB)=m$ (see examples in Section \ref{examples}), then we have following corollary.

\begin{corollary}   \label{col: Main Theorem}

Let $\vx_1,\ldots,\vx_m$ be i.i.d. centered sub-Gaussian random vectors with positive definite covariance matrix $\vSigma \in \R^{n \times n}$ and $\bm{X}=[\vx_1,\ldots,\vx_m] \in \R^{n \times m}$. Consider the correlated sample covariance matrix estimator $\hat{\vSigma} = \vX \vB \vX^T/m,$ where  $\vB \in \R ^{m \times m}$ is an arbitrary matrix satisfying $\tr(\vB)=m$. Suppose that there exists $K\ge1$ such that
\begin{equation*}
\norm{\ip{\vx_i}{\vy}}_{\psi_2} \le K \norm{\ip{\vx_i}{\vy}}_{L^2}, \forall \,  \vy \in \R^n,\, i=1,\ldots,m.
\end{equation*}
Then for any $\delta \geq 0$, the estimator satisfies
	\begin{equation*}
	\norm{\hat{\vSigma}-\vSigma} \le  CK^2 \(\frac{\sqrt{n+\delta} \norm{\vB}_F + (n+\delta) \norm{\vB}}{m}\)\norm{\vSigma}
	\end{equation*}
	with probability at least $1-2\exp(-\delta)$.Furthermore,
	\begin{equation*}
	\E \norm{\hat{\vSigma}-\vSigma} \le CK^2 \(\frac{\sqrt{n} \norm{\vB}_F + n \norm{\vB}}{m}\)\norm{\vSigma}.
	\end{equation*}
\end{corollary}

Next, we are going to present a variant of Theorem \ref{thm: General} in the complex domain for $\vB\in \C^{m \times m}$. Combining Theorem \ref{thm: wishart_SG2} and the triangle inequality as the proof of Theorem \ref{thm: General}, we can obtain the following results.

\begin{theorem}[Complex case] \label{thm: General2}
	Consider a complex vector $\vx \in \C^n$ whose  real part ${\rm{Re}}(\vx)$ and imaginary part ${\rm{Im}}(\vx)$ are i.i.d. centered sub-Gaussian vectors. Let the covariance matrix of $\vx$ be positive definite， denoted by $\vSigma=\E \vx \vx^H$. Suppose that there exists $K\ge1$ such that
	\begin{equation*}
	\norm{\ip{\left[
			\begin{array}{c}
			{\rm{Re}} (\vx)       \\
			{\rm{Im}} (\vx)          \\
			\end{array}
			\right]}{\vy}}_{\psi_2} \le 2K \norm{\ip{\left[
			\begin{array}{c}
			{\rm{Re}} (\vx)       \\
			{\rm{Im}} (\vx)          \\
			\end{array}
			\right]}{\vy}}_{L^2}
	\end{equation*}
	for any $\vy \in \R^{2n}$.
	Let  vectors $\vx_1,\ldots,\vx_m \in \C^n$ be independent copies of $\vx$ and $\bm{X}=[\vx_1,\ldots,\vx_m] \in \C^{n \times m}$. Consider the correlated sample covariance matrix $\hat{\vSigma}=\vX \vB \vX^H/m$, where $\vB \in \C^{m \times m}$ is a fixed matrix.
	{Then for any $\delta \geq 0$, we have
		\begin{multline*}
	\norm{\hat{\vSigma}-\vSigma} \le \left| \frac{\tr(\vB)}{m}-1 \right|  ||\vSigma||
	\\ + CK^2 \(\frac{\sqrt{n+\delta} \norm{\vB}_F + (n+\delta) \norm{\vB}}{m}\)\norm{\vSigma}
	\end{multline*}
	with probability at least $1- {c}\exp(-\delta)$.} Furthermore,
	\begin{multline*}
	\E \norm{\hat{\vSigma}-\vSigma} \le  \left| \frac{\tr(\vB)}{m}-1 \right|  ||\vSigma|| \\
	+ CK^2 \(\frac{\sqrt{n} \norm{\vB}_F + n  \norm{\vB}}{m}\)\norm{\vSigma}.
	\end{multline*}
\end{theorem}

 \subsection{Examples}\label{examples}
 In this subsection, we provide three special correlation patterns to illustrate our theoretical results.
 \begin{example}[Independent sub-Gaussian samples] The independent samples imply $\vB=\vI_m$. It follows from Corollary \ref{col: Main Theorem} that
	\begin{equation*}
	\E \norm{\hat{\vSigma}-\vSigma} \le CK^2 \(\sqrt{\frac{n}{m}} + \frac{n}{m}\)\norm{\vSigma}.
	\end{equation*}
	 Therefore, we require $m=O(n)$ independent sub-Gaussian samples to accurately estimate the covariance matrix, which is consistent with Proposition \ref{pp: sg}.
 \end{example}

 \begin{example}[Partially correlated sub-Gaussian samples with Hermitian shape parameter] A popular model for the  correlation pattern is a class of Hermitian Toeplitz matrices, i.e.,
 	\begin{equation*}
 	\vB=\left[
 	\begin{array}{cccccc}
 	1      & \omega      & \cdots & \omega^{m-1}  \\
 	\bar{\omega}      & 1      & \ddots & \vdots   \\
 	\vdots & \ddots & \ddots & \omega        \\
 	\bar{\omega}^{m-1}& \cdots & \bar{\omega}      & 1        \\
 	\end{array}
 	\right]\doteq\vT(\omega)
 	\end{equation*}
 	with $\omega \in \C$ and $0 < |\omega| < 1$. When $w$ is a real number, a typical application is the lagged correlation between the returns in portfolio optimization \cite{burda2011applying}, which meets this model by setting $w=\exp(-1/\tau)$. Here $\tau \in \R$ is the characteristic time.
 	
 	By using Gershgorin circle theorem \cite[Theorem 7.2.1]{golub2012matrix}, we obtain
 	$$
 	\norm{\vT(\omega)}\le1+\sum_{k=1}^{\infty}|\omega|^k+\sum_{k=1}^{\infty}|\bar{\omega}|^k\le 1+\frac{2|w|}{1-|w|}=\frac{1+|w|}{1-|w|},
 	$$
 	for $0 < |\omega| < 1$. And the Frobenius norm of $\vT(\omega)$ is
 	\begin{align*}
 	\norm{\vT(\omega)}_F^2
 	&=m+2\cdot \sum_{k=1}^{m-1}(m-k)|w|^{2k}\\
 	&=\frac{m\left(1+|\omega|^{2}\right)}{1-|\omega|^{2}}+\frac{2 |\omega|^{2}\left(|\omega|^{2 m}-1\right)}{\left(1-|\omega|^{2}\right)^{2}}\\
 	&\le \frac{m\left(1+|\omega|^{2}\right)}{1-|\omega|^{2}}.
 	\end{align*}
 	Note that $\mbox{tr}(\vT(\omega))=m$. By Theorem \ref{thm: General2}, we have
 	\begin{equation*} \label{leq: T}
 	\mathbb{E}\norm{\hat{\vSigma}-\vSigma} \le CK^2 \( \sqrt{\frac{1+|\omega|^2}{1-|\omega|^2} \cdot \frac{n}{m}} +  \frac{1+|\omega|}{1-|\omega|} \cdot\frac{n}{m} \) \norm{\vSigma}.
 	\end{equation*}
 	 	
The above results reveal that in this case, $m=O(n)$ samples are sufficient to accurately estimate the covariance matrix from correlated sub-Gaussian samples. In contrast with the independent case, this correlated case requires more samples to achieve the same estimation accuracy since we have an additional multiplier coefficient $((1+|\omega|)/(1-|\omega|)>1)$ in the error bound. This is consistent with our intuition. Another important conclusion is that the larger the parameter $|w|$ is, the more correlated samples we require.
 \end{example}

 \begin{example}[Partially correlated sub-Gaussian samples with non-Hermitian shape parameter]
 	In this example, we consider non-Hermitian shape parameter. The non-Hermitian shape parameter $\vP(c,\bm{\Theta})$ is constructed as follows. Let $0<c<1$ be a real number and $\bm{\Theta} \in \R^{m \times m}$ be a square matrix. The $(a,b)$-th entry of $\vP(c,\Phi)$ is
	$$
	[\vP(c,\bm{\Theta})]_{ab} =(c\,e^{j\Theta_{ab}})^{|a-b|} \in \C,\,a,b=1,\ldots,m,
	$$	
	where the entries $\{\Theta_{ab}\}$ can be arbitrary numbers in the range $[0,2\pi)$ and $j = \sqrt{-1}$.

	It then follows from Example 2 that
	\begin{equation*} \label{leq: T}
	\mathbb{E}\norm{\hat{\vSigma}-\vSigma} \le CK^2 \( \sqrt{\frac{1+c^2}{1-c^2} \cdot \frac{n}{m}} +  \frac{1+c}{1-c} \cdot\frac{n}{m} \) \norm{\vSigma}.
	\end{equation*}
    The example illustrates that for this non-Hermitian correlation pattern, $O(n)$ samples are enough to approximate the covariance matrix accurately.
\end{example}


\begin{figure}[!t]
	\centering
	\includegraphics[width=3in]{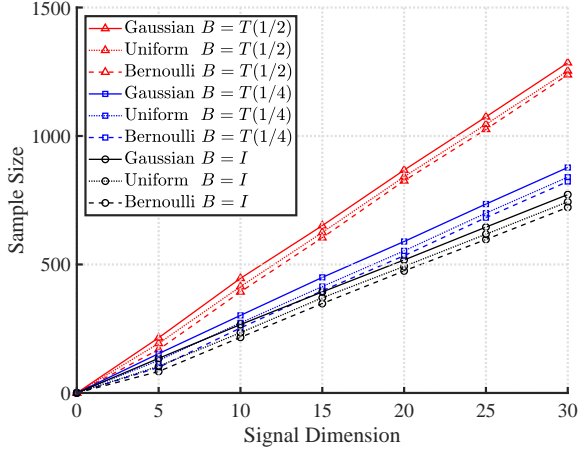}
	\caption{Sample size v.s. signal dimension for Gaussian, uniform and Bernoulli real random samples under three real correlated models.}
	\label{Fig1}
\end{figure}

\begin{figure}[!t]
	\centering
	\includegraphics[width=3in]{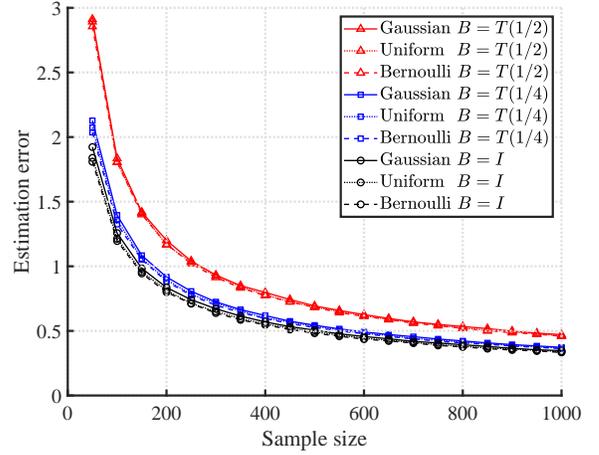}
	\caption{Convergence rate for Gaussian, uniform and Bernoulli real random samples under three real correlated models.}
	\label{Fig2}
\end{figure}

\begin{figure}[!t]
	\centering
	\includegraphics[width=3in]{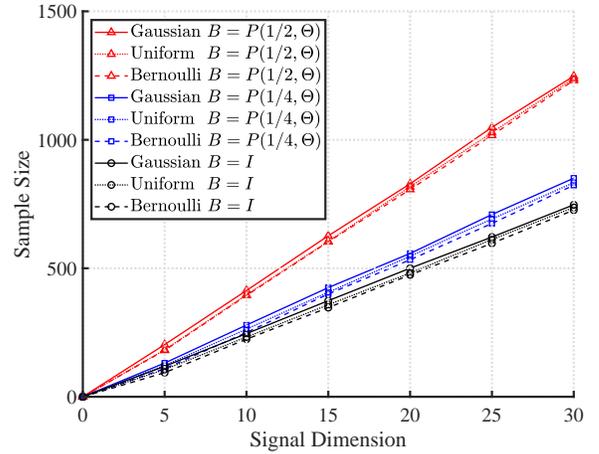}
	\caption{Sample size v.s. signal dimension for Gaussian, uniform and Bernoulli complex random samples under three complex correlated models.}
	\label{Fig3}
\end{figure}

\section{Numerical Simulations} \label{sec: Simulation}

In this section, we present some simulations to verify the theoretical results.

Let $\vX$ be a random matrix with dimension ${n \times m}$ whose entries are i.i.d. with zero mean and unit variance. Let $\vB$ denote the correlation pattern (shape parameter) with dimension ${m \times m}$.

In the first simulation, we show the relationship between sample size and signal dimension for three kinds of real random samples under three kinds of real correlation patterns. The three kinds of random samples include standard Gaussian, uniform, and symmetric Bernoulli random samples. The correlation patterns are: 1) $\vB=\vI$; 2) $\vB=\vT(1/4)$; 3) $\vB=\vT(1/2)$. The tolerance is set as $\eta=0.2$ and the signal dimension $n$ increases from 0 to 30. For each signal dimension, $500$ Monte-Carlo trials are made to calculate the average of the minimum sample size $m$ that satisfies the normalized mean square error condition
$$
\frac{||\hat\vSigma - \vSigma||_F}{ ||\vSigma||_F} \le \eta.
$$

The results are shown in Fig. \ref{Fig1}. From the simulation, we know for all cases, the sample size is a linear function of the signal dimension, which means that $O(n)$ samples are enough to estimate the covariance matrix. Besides, when the model gets more correlated, we need more samples to achieve the given precision for the same kind of random samples. The simulation results agree with the theoretical results shown in Corollary \ref{col: Main Theorem}.

In the second simulation, we consider the convergence curve for Gaussian, uniform, and Bernoulli real random samples under the above three types of correlation patterns.  We set $n=30$ and increase $m$ from 50 to 1000 with step 50. For each sample size $m$, 500 Monte Carlo trials are performed to average the estimation error $||\hat{\vSigma}-\vSigma||$. The results are presented in Fig. \ref{Fig2}. From the figure, we can see the three random samples have similar convergence curve. For the same kind of random samples under different correlated models, Fig. \ref{Fig2} shows that the more correlated the model is, the worse the convergence curve is. The results coincide with our theory (Theorem \ref{thm: General}).

In the third simulation, we give the relationship between signal dimension and sample size for complex random samples under general complex correlation patterns.
The complex random samples include Gaussian, uniform, and Bernoulli random samples, whose real part and imaginary part are i.i.d.. The correlation patterns are: 1) $\vB=\vI$; 2) $\vB=\vP(1/4,\bm{\Theta})$; 3) $\vB=\vP(1/2,\bm{\Theta})$. The entries of $\bm{\Theta}$ are generated randomly from $[0,2\pi)$. The other simulation settings are the same as the first simulation. The results are given in the Fig. \ref{Fig3}. Similar to Fig. \ref{Fig1}, the requierd sample size is linear with the signal dimension, which can be explained by Theorem \ref{thm: General2}. Furthermore, with the increase of $c$, we require more correlated samples to achieve the same precision.

\section{Conclusion and future work} \label{sec: Conclusion}
In this paper, we have analyzed the problem of covariance matrix estimation from correlated sub-Gaussian samples. The non-asymptotic error bounds have been established for this problem in both tail and expectation forms. These error bounds are determined by the sample size $m$, the signal dimension $n$, and the shape parameter $\vB$. In some applications of interest, where the shape parameter $\vB$ satisfies $\tr(\vB)=m$, $||\vB||_{F}=O(m^{1/2})$ and $||\vB||=O(1)$, our results indicate that $O(n)$ correlated sub-Gaussian samples to estimate the covariance matrix accurately. An extension of the theory to complex domain has been made to meet the requirement for signal processing applications.

There are some interesting problems stemming from this work. An important problem is to consider covariance matrix estimation from correlated heavy-tailed samples. Another problem is to extend the work to other estimators, such as structured estimators, regularized estimators and so on.


%

\appendices

\section{Proof of Theorem \ref{thm: wishart_SG1}} \label{appdix: Proof_Thm1}
Without loss of generality, we assume $\vSigma= \vI$, otherwise we can use $\vSigma^{-1/2} \vX$ instead of ${\vX}$ to verify the general case. Thus $\{\vx_i\}_{i=1}^{m}$ are i.i.d. centered isotropic sub-Gaussian random vectors and the condition \eqref{eq: psi_2_norm_condition2} becomes $\norm{\vx_i}_{\psi_2} \le K$ for any $i \in \{1,\ldots,m\}$.

For clarity, the proof is divided into several steps.
\begin{itemize}
	\item[1)] {\bf{Approximation}}. It follows from Fact \ref{lm: covering} (by choosing $\e=1/4$) that
	\begin{align*}
		&\norm{\hat{\vSigma}-\E\hat{\vSigma}} \\
		&\le 2 \sup \limits_{\vu \in \mathcal{N},\vv \in \mathcal{M}}
		\ip{ \(\hat{\vSigma}-\E\hat{\vSigma}\)\vu}{\vv}\\
		&\le 2 \sup \limits_{\vu \in \mathcal{N},\vv \in \mathcal{M}}
		\left|\ip{ \hat{\vSigma}\vu}{\vv}-\ip{\(\E\hat{\vSigma}\)\vu}{\vv}\right|\\
		&= 2 \sup \limits_{\vu \in \mathcal{N},\vv \in \mathcal{M}}
		\left|\ip{ \hat{\vSigma}\vu}{\vv}-\E\ip{\hat{\vSigma}\vu}{\vv}\right|\\
		&=\frac{2}{m} \sup \limits_{\vu \in \mathcal{N},\vv \in \mathcal{M}}
		\left|\ip{\vB \vX^T \vu}{\vX^T\vv}-\E\ip{\vB \vX^T \vu}{\vX^T\vv}\right|,	 	
	\end{align*}
	where $\mathcal{N}$ is a ${1}/{4}$-net of $\mathbb{S}^{n-1}$ with $|\mathcal{N}| \le 9^n$ and $\mathcal{M}$ is also a ${1}/{4}$-net of $\mathbb{S}^{n-1}$ with $|\mathcal{M}| \le 9^n$. Define $\vz^{\vu}=\vX^T\vu=[\vx_1^T\vu,\ldots,\vx_m^T\vu]^T$ and $\vz^{\vv}=\vX^T\vv=[\vx_1^T\vv,\ldots,\vx_m^T\vv]^T$.
	Then we have
	\begin{multline*}
		\P(\norm{\hat{\vSigma}-\E\hat{\vSigma}}  \ge t)
		\\ \le \P\( \sup \limits_{\vu \in \mathcal{N},\vv \in \mathcal{M}}
		\left|\ip{\vB \vz^{\vu}}{\vz^{\vv}}-\E\ip{\vB \vz^{\vu}}{\vz^{\vv}}\right|	 	 \ge \frac{mt}{2}\).
	\end{multline*}
	\item[2)] {\bf{Concentration}}. Fixing  $\vu \in \mathcal{N}$ and $\vv \in \mathcal{M}$, we will establish the tail bound
	$$\P\(
	\left|\ip{\vB \vz^{\vu}}{\vz^{\vv}}-\E\ip{\vB \vz^{\vu}}{\vz^{\vv}}\right| \ge \frac{mt}{2}\).
	$$
	Observe that
    \begin{align*}
		\ip{\vB \vz^{\vu}}{\vz^{\vv}}&=\ip{\vB \vX^T \vu}{\vX^T\vv}\\
		&=\tr (\vu^T \vX \vB^T \vX^T \vv)\\
		&=\tr ( \vB^T \vX^T \vv \vu^T \vX)\\
		&=\[\text{vec}(\vX \vB)\]^T \text{vec}(\vv \vu^T \vX),
	\end{align*}
   where the last two equalities follow from $\tr(\vA\vB) = \tr(\vB\vA)$ and $\tr(\vA\vB) = \text{vec}(\vA^T)^T\text{vec}(\vB)$ respectively. Thus we have
	\begin{align*}
	&\ip{\vB \vz^{\vu}}{\vz^{\vv}}=\[\text{vec}(\vI_n\vX \vB)\]^T \text{vec}(\vv \vu^T \vX \vI_m)\\
	&=\[\(\vB^T \otimes \vI_n\)\text{vec}(\vX)\]^T\[\(\vI_m \otimes \vv \vu^T\)\text{vec}(\vX)\]\\
	&=\text{vec}(\vX)^T (\vB \otimes \vI_n)\(\vI_m \otimes \vv \vu^T\)\text{vec}(\vX)\\
	&=\text{vec}(\vX)^T \(\vB \otimes \vv \vu^T\)\text{vec}(\vX),
	\end{align*}
	where the second equality holds because $\text{vec}(\vA \vX \vB)=\( \vB^T \otimes \vA\) \text{vec}(\vX)$, the third equality follows from $(\vA \otimes \vB)^T = (\vA^T \otimes \vB^T)$, and the fourth equality uses the fact $(\vA\otimes\vB)(\vC \otimes \vD )=(\vA\vC) \otimes(\vB\vD)$.
	
	In order to use Fact \ref{lm: HW}, we need to calculate $\norm{\vB \otimes \vv \vu^T}$ and
	$\norm{\vB \otimes \vv \vu^T}_F$ first. According to \cite[Theorem 4.2.15]{roger1994topics}, we have $\norm{\vA \otimes \vB} =\norm{\vA}\norm{\vB}$ and $\norm{\vA \otimes \vB}_F =\norm{\vA}_F \norm{\vB}_F$. Since $\vv, \vu \in \S^{n-1}$, we obtain
	$$\norm{\vB \otimes \vv \vu^T}= \norm{\vB} \norm{\vv \vu^T}=\norm{\vB} $$ and $$\norm{\vB \otimes \vv \vu^T}_F= \norm{\vB}_F\norm{\vv \vu^T}_F= \norm{\vB}_F.$$
	In addition, from the definition of sub-Gaussian vectors, we know that each entry of $\text{vec}(\vX)$ is a sub-Gaussian variable with sub-Gaussian norm less than or equal to $K$.
	It then follows from Fact \ref{lm: HW} that
	\begin{multline*}
	\P\(
	\left|\ip{\vB \vz^{\vu}}{\vz^{\vv}}-\E\ip{\vB \vz^{\vu}}{\vz^{\vv}}\right| \ge \frac{mt}{2}\) \\
	\le 2 \exp\[-c \min \left\{\frac{m^2t^2}{K^4 \norm{\vB}_F^2}, \frac{mt}{ K^2\norm{\vB}}\right\}\].
	\end{multline*}
	\item[3)] {\bf{Tail bound}}. Taking union bound for all $\vu \in \mathcal{N}$ and $\vv \in \mathcal{M}$ yields
	\begin{multline*}
	\P\( \sup \limits_{\vu \in \mathcal{N}}
	\left|\ip{\vB \vz^{\vu}}{\vz^{\vv}}-\E\ip{\vB \vz^{\vu}}{\vz^{\vv}}\right| \ge \frac{mt}{2} \) \\
	\le 9^{2n} \cdot 2 \exp\[-c \min \left\{\frac{m^2t^2}{K^4 \norm{\vB}_F^2}, \frac{mt}{ K^2\norm{\vB}}\right\}\].
	\end{multline*}
	
	Assigning
	$$
	    t= CK^2 \(\frac{ \sqrt{n+\delta}\norm{\vB}_F+(n+\delta) \norm{\vB}}{m}\)\doteq t_1,
	$$
     we obtain
	\begin{multline*}
		\P\( \sup \limits_{\vu \in \mathcal{N}}
		\left|\ip{\vB \vz^{\vu}}{\vz^{\vu}}-\E\ip{\vB \vz^{\vu}}{\vz^{\vu}}\right| \ge \frac{mt_1}{2}\) \\
		\le 2 \exp\(- \delta\)
	\end{multline*}
   for a large enough constant $C$.
   Therefore, we show that
	\begin{multline*}
		\norm{\hat{\vSigma}-\E\hat{\vSigma}}  \\ \ge CK^2 \(\frac{ \sqrt{n+\delta}\norm{\vB}_F+(n+\delta) \norm{\vB}}{m}\)
	\end{multline*}
   holds with probability at most $2 \exp\(- \delta\)$.
   In particular, if $s^2=\delta \ge n$, the following event
    $$
	\norm{\hat{\vSigma}-\E\hat{\vSigma}}  \ge C' K^2 \(\frac{ s \norm{\vB}_F+s^2 \norm{\vB}}{m}\)\doteq t_2
	$$
	holds with probability at most $ 2 \exp\(- s^2\)$, which is used to establish the expectation bound.
	
	\item[4)] {\bf{Expectation bound}}. Note that
	\begin{align*}
		&\E\norm{\hat{\vSigma}-\E\hat{\vSigma} }\\
		&= \int_0^\infty {\mathbb{P}\left(\norm{\hat{\vSigma}-\E\hat{\vSigma}} \ge t\right) {\rm{d}} t}\\
		&=\frac{C'K^2}{m} \int_0^\infty \mathbb{P}\left(\norm{\hat{\vSigma}-\E\hat{\vSigma}} \ge t_2\right)\\
		& \qquad \qquad \qquad \qquad \qquad \quad \( \norm{\vB}_F+2s \norm{\vB}\) {\rm{d}} s\\
		&\le \frac{C'K^2}{m} \int_0^{\sqrt{n}} 1 \cdot \( \norm{\vB}_F+2s \norm{\vB}\) {\rm{d}} s \\
		&~~~ + \frac{{2}C'K^2}{m}\int_{\sqrt{n}}^\infty \exp \(-s^2\)\( \norm{\vB}_F+2s \norm{\vB}\)  {\rm{d}} s\\
		&\le C''K^2 \frac{ \sqrt{n}\norm{\vB}_F+n \norm{\vB}}{m},
	\end{align*}
	where the first equality is due to the integral identity, in the second inequality we have let $t = C' K^2 \(\frac{ s \norm{\vB}_F+s^2 \norm{\vB}}{m}\)$, and the last inequality holds by choosing a large enough constant $C''$. Thus we complete the proof.

\end{itemize}

\section{Proof of Theorem \ref{thm: wishart_SG2}} \label{appdix: Proof_Thm3}

	In order to extend Theorem \ref{thm: wishart_SG1} from real domain to complex domain, we require the complex version of Definitions 1-4 and Facts 1-2. It is not hard to check that Definitions 1-4 and Fact 1 can be easily extended the complex case, see e.g., \cite{tao2012topics}.
	
	We then extend Fact \ref{lm: HW} to complex domain by following the technique from the proof of Theorem 1.4 in \cite{vu2015random}.
	
	\begin{lemma} \label{lm: HW2} Assume that $\vx \in \C^m$ has i.i.d. real and imaginary parts. Its entries $\{x_i\}$ are independent centered sub-Gaussian variables with $\norm{{\rm{Re}} (x_i)}_{\psi_2}\le K$ and $\norm{{\rm{Im}} (x_i)}_{\psi_2}\le K$ for $i=1,\ldots,m$. Let $\vB \in \C^{m \times m}$ be a fixed matrix. Then for any $t\geq0$,
		\begin{multline*}
		\P\(\left|\vx^H\vB \vx - \E \[\vx^H\vB \vx\]\right| \ge t\)
		\\   \le
		C \exp\[-c\, \min\(\frac{t^2}{K^4 \norm{\vB}_F^2}, \frac{t}{K^2\norm{\vB}}\) \].
		\end{multline*}
		where $C$ and $c$ are absolute constants.
	\end{lemma}
	\begin{IEEEproof} 
    See Appendix \ref{appdix: Proof_Lemmacomplex}.
    \end{IEEEproof}
	
	We are now in position to prove of Theorem \ref{thm: wishart_SG2}.

	Since  ${\rm{Re}} (\vx_i)$ and ${\rm{Im}} (\vx_i)$ are i.i.d.,  we have $ \E {\rm{Re}} (\vx_i) {\rm{Re}} ^T(\vx_i)= \E {\rm{Im}} (\vx_i) {\rm{Im}} ^T(\vx_i)=\vSigma/2$.
	As before, we assume $\vSigma= \vI$, otherwise we can use $\vSigma^{-1/2} \vX$ instead of ${\vX}$ to verify the general case.  In this case, the condition \eqref{eq: psi_2_norm_condition_complex} becomes
	\begin{equation*}
	\norm{\left[
		\begin{array}{c}
		{\rm{Re}} (\vx_i)       \\
		{\rm{Im}} (\vx_i)       \\
		\end{array}
		\right]}_{\psi_2} \le K.
	\end{equation*}
	
	\begin{itemize}
		\item[1)] {\bf{Approximation}}. By Fact \ref{lm: covering}, we get
		\begin{multline*}
		\P(\norm{\hat{\vSigma}-\E\hat{\vSigma}}  \ge t)
		\\ \le \P\( \sup \limits_{\vu \in \mathcal{N},\vv \in \mathcal{M}}
		\left|\ip{\vB \vz^{\vu}}{\vz^{\vv}}-\E\ip{\vB \vz^{\vu}}{\vz^{\vv}}\right|	 	 \ge \frac{mt}{2}\).
		\end{multline*}
		where $\mathcal{N}$ is a ${1}/{4}$-net of $\mathbb{S}^{n-1}$ with $|\mathcal{N}| \le 9^n$, $\mathcal{M}$ is also a ${1}/{4}$-net of $\mathbb{S}^{n-1}$ with $|\mathcal{M}| \le 9^n$, and $\S^{n-1}$ denotes the unit sphere in $\C^n$. Here, $\vz^{\vu}=\vX^H\vu=[\vx_1^H\vu,\ldots,\vx_m^H\vu]^T$, $\vz^{\vv}=\vX^H\vu=[\vx_1^H\vv,\ldots,\vx_m^H\vv]^T$.
		
		\item[2)] {\bf{Concentration}}. Fix $\vu \in \mathcal{N}$ and $v \in \mathcal{M}$. Similar to the real case, we have
		\begin{align*}
		&\ip{\vB \vz^{\vu}}{\vz^{\vv}}=\ip{\vB \vX^H \vu}{\vX^H\vv}\\
		&=\tr (  \vu^H \vX \vB^H \vX^H \vv )\\
		&=\tr (  \vB^H \vX^H \vv \vu^H \vX  )\\
		&=\text{vec}(\vI_n \vX \vB)^H \text{vec}(\vv \vu^H \vX \vI_m)\\
		&=\[\(\vB^T \otimes \vI_n\)\text{vec}(\vX)\]^H\[\(\vI_m \otimes \vv \vu^H\)\text{vec}(\vX)\]\\
		&=\text{vec}(\vX)^H \(\bar{\vB} \otimes \vv \vu^H\)\text{vec}(\vX),
		\end{align*}
		where the fourth line follows from $\mbox{tr}(\vX\vY)=\mbox{vec}(\vX^H)^H\mbox{vec}(\vY)$, the last inequality holds because $(\vA \otimes \vB)^H = (\vA^H \otimes \vB^H)$ and $(\vA\otimes\vB)(\vC \otimes \vD )=(\vA\vC) \otimes(\vB\vD)$, and $\bar{\vB}$ denotes the complex conjugate of $\vB$.
	
	    Notice that
		$\norm{\bar{\vB} \otimes \vv \vu^H}= \norm{\vB}$ and $\norm{\bar{\vB} \otimes \vv \vu^H}_F= \norm{\vB}_F.$
		Since each entry of $\text{vec}(\vX)$ is independent sub-Gaussian and the sub-Gaussian norm of real part and imaginary part is less than $K$, using Fact \ref{lm: HW2} yields
		\begin{multline*}
		\P\(
		\left|\ip{\vB \vz^{\vu}}{\vz^{\vu}}-\E\ip{\vB \vz^{\vu}}{\vz^{\vu}}\right| \ge \frac{mt}{2}\) \\
		\le
		C \exp\[-c\, \min\(\frac{m^2t^2}{K^4 \norm{\vB}_F^2},\frac{mt}{K^2\norm{\vB}}\) \].
		\end{multline*}
		\item[3)] {\bf{Tail bound and expectation bound}}. Just like the proof in Appendix \ref{appdix: Proof_Thm1}, taking union bound and integrating the probability enable us to get the final results.	The proof is very similar, so we ignore it here.
	\end{itemize}

\section{Proof of Lemma \ref{lm: HW2}} \label{appdix: Proof_Lemmacomplex}

We first show that it is sufficient to establish the lemma for positive semidefinite $\vB$.

        Let $y = \vx^H\vB\vx \in \C$. Its complex conjugate is $\bar{y}=\vx^H\vB^H\vx$. Then we have $y+\bar{y}=\vx^H(\vB+\vB^H)\vx$ and $y-\bar{y}=\vx^H(\vB-\vB^H)\vx$. Note that
		\begin{align*}
		y-\E y&=\frac{1}{2}\[(y+\bar{y})-\E(y+\bar{y})\]+\frac{1}{2}\[(y-\bar{y})-\E(y-\bar{y})\]\\
		&=\frac{1}{2}\left\{\vx^H(\vB+\vB^H)\vx-\E\[\vx^H(\vB+\vB^H)\vx\]\right\}\\
		&\quad+\frac{1}{2}\left\{\vx^H(\vB-\vB^H)\vx-\E\[\vx^H(\vB-\vB^H)\vx\]\right\}.
		\end{align*}
		Thus we have
		\begin{align}\label{equation1}
		&\P\(\left|\vx^H\vB \vx - \E \[\vx^H\vB \vx\]\right| \ge t\) \le\\ \notag
		&\quad \P\(|\vx^H(\vB+\vB^H)\vx-\E\[\vx^H(\vB+\vB^H)\vx\]|\ge t\)\\ \notag
		&\quad +\P\(|\vx^H(j\vB-j\vB^H)\vx-\E\[\vx^H(j\vB-j\vB^H)\vx\]| \ge t\),
		\end{align}
		where $j=\sqrt{-1}$ denotes the imaginary unit.

        Observe that $\norm{\vB+\vB^H}_F$ and $\norm{j\vB-j\vB^H}_F$ have the order of $\norm{\vB}_F$ (i.e., $O(\norm{\vB}_F)$), and $\norm{\vB+\vB^H}$ and $\norm{j\vB-j\vB^H}$ have the order of $\norm{\vB}$ (i.e., $O(\norm{\vB})$). Since $\vB+\vB^H$ and $j(\vB-\vB^H)$ are Hermitian matrices, it is enough to prove the lemma for Hermitian $\vB$.
		
		It is well known that any Hermitian matrix $\vB$ can be decomposed as $\vB=\vB_1-\vB_2$, where $\vB_1$ and $\vB_2$ are positive semi-definite matrices with $\norm{\vB_1},\norm{\vB_2} \le \norm{\vB}$ and $\norm{\vB_1}_F,\norm{\vB_2}_F \le \norm{\vB}_F$. Indeed, $\vB_1$ can be constructed by using the positive eigenvalues and corresponding eigenvectors of $\vB$ while  $\vB_2$ can be constructed by using the negative eigenvalues and corresponding eigenvectors of $\vB$. Then we obtain
		\begin{align}\label{equation2}
		&\P\(\left|\vx^H\vB \vx - \E \[\vx^H\vB \vx\]\right| \ge t\) \le\\ \notag
		&\qquad \qquad\P\(|\vx^H\vB_1\vx-\E\[\vx^H\vB_1 \vx\]|\ge t/2\)\\ \notag
		& \qquad \qquad+\P\(|\vx^H \vB_2 \vx-\E\[\vx^H\vB_2\vx\]| \ge t/2\).
		\end{align}
		Therefore, it suffices to prove the lemma for  positive semi-definite $\vB$.
		
		Without loss of generality, we assume $\vB$ is a positive semi-definite matrix and decompose it as $\vB=\vLambda\vLambda^H$. Then have
		\begin{multline*}
		\P\(|\vx^H\vB \vx - \E\vx^H\vB \vx| \ge t\)\\=\P\(\left|\norm{\vLambda^H\vx}_2^2 - \E \norm{\vLambda^H\vx}_2^2 \right| \ge t\).
		\end{multline*}
		Define
		\begin{equation*}
		\vA=\left[
		\begin{array}{cc}
		{\rm{Re}} (\vLambda)       &    -{\rm{Im}} (\vLambda)   \\
		{\rm{Im}} (\vLambda)      &  {\rm{Re}} (\vLambda)       \\
		\end{array}
		\right] \text{~and~} \vz=\left[
		\begin{array}{c}
		{\rm{Re}} (\vx)       \\
		{\rm{Im}} (\vx)          \\
		\end{array}
		\right],
		\end{equation*}
		where $\vA \in \R^{2m \times 2m} $ and $\vz  \in \R^{2m}$.
		Note that $\norm{\vA^T \vz}_2=\norm{\vLambda^H\vx}_2$. Thus we can change the probability from complex domain to real domain
		\begin{multline*}
		\P\(\left|\norm{\vLambda^H\vx}_2^2 - \E \norm{\vLambda^H\vx}_2^2 \right| \ge t\)\\
		= \P\(\left|\norm{\vA^T \vz}_2^2 - \E \norm{\vA^T \vz}_2^2 \right| \ge t\).
		\end{multline*}
		It follows from Fact \ref{lm: HW} that
		\begin{multline*}
		\P\(\left|\norm{\vA^T \vz}_2^2 - \E \norm{\vA^T \vz}_2^2 \right| \ge t\)\\
		\le 2 \exp\[-c\, \min\(\frac{t^2}{K^4\norm{\vA\vA^T}_{F}^2}, \frac{t}{K^2\norm{\vA\vA^T}}\) \].
		\end{multline*}
		
		Note that $\norm{\vA\vA^T}=\norm{\vA}^2=\norm{\vLambda}^2=\norm{\vB}$ and  $\norm{\vA\vA^T}_F^2=2\norm{\vB}_F^2$. Thus we have
		\begin{multline}\label{equation3}
		\P\(|\vx^H\vB \vx - \E\vx^H\vB \vx| \ge t\)\\
		\le 2 \exp\[-c\, \min\(\frac{t^2}{K^4 \norm{\vB}_F^2},\frac{t}{K^2\norm{\vB}} \) \].
		\end{multline}
		
    For general $\vB$, combining \eqref{equation1}, \eqref{equation2}, \eqref{equation3} and taking union bound yields the desired result.


\ifCLASSOPTIONcaptionsoff
  \newpage
\fi



\end{document}